\documentclass[letterpaper, 10 pt, conference]{ieeeconf}   
\IEEEoverridecommandlockouts                    
\let\labelindent\relax
\usepackage{enumitem}

\usepackage{mathtools,amsthm}

\usepackage{graphics} 
\usepackage{epsfig}
\usepackage{times} 
\usepackage{amsmath,amssymb,amsfonts}
\usepackage{mathrsfs}
\usepackage{algorithmic}
\usepackage{url}
\usepackage{graphicx}
\usepackage{textcomp}
\usepackage{varioref}
\usepackage{import}
\usepackage{framed,xcolor}
\usepackage{color}
\usepackage{comment}
\usepackage{multirow}
\usepackage{epstopdf}   
\usepackage{psfrag}
\usepackage{breqn}
\usepackage{float}
\usepackage{mathtools}
\usepackage{booktabs}
\usepackage{soul}
\usepackage{amsbsy}
\usepackage[bottom]{footmisc}
\usepackage{lineno}
\usepackage{cleveref}
\usepackage{microtype}
\usepackage{siunitx}
\usepackage{comment}
\usepackage{multicol}

\definecolor{arash}{rgb}{0.8,0.8,1}
\definecolor{seb}{rgb}{0.8,1,0.8}
\definecolor{seb2}{rgb}{0.5,.5,1}
\definecolor{arash2}{rgb}{0,.5,0}
\definecolor{wenqi}{rgb}{1,.75,0.79}
\definecolor{wenqi2}{rgb}{1,.75,0.79}
\graphicspath{{Figures/}}

\graphicspath{{Figures/}}
\newcommand{\vect}[1]{\ensuremath{\boldsymbol{\mathrm{#1}}}}
\usepackage{graphicx}
\usepackage{tabularx,booktabs}
\usepackage{color}
\usepackage{multicol}
\usepackage{amsmath,amssymb}
\usepackage{amsfonts}
\usepackage{textcomp}
\usepackage{psfrag}
\usepackage{multimedia}
\usepackage{fancybox}
\usepackage{comment}
\usepackage{soul}
\makeatletter
\newcommand{\biggg}{\bBigg@{1.6}}  

\makeatother

\definecolor{seb}{rgb}{0.8,1,0.8}
\definecolor{arash}{rgb}{0.8,0.8,1}

\newcounter{lastnote}

\title{\LARGE \bf
Optimal Management of the Peak Power Penalty for Smart Grids Using MPC-based Reinforcement Learning}
\author{Wenqi Cai, Hossein N. Esfahani, Arash B. Kordabad, and Sébastien Gros
\thanks{The authors are with Department of Engineering Cybernetics, Norwegian University of Science and Technology (NTNU), Trondheim, Norway. E-mail:{\tt\small\{wenqi.cai,hossein.n.esfahani,arash.b.korda\newline bad,sebastien.gros\}@ntnu.no}}  
}
\begin{document}
\maketitle
\thispagestyle{empty}
\pagestyle{empty}
%%%%%%%%%%%%%%%%%%%%%%%%%%%%%%%%%%%%%%%%%%%%%%%%%%%%%%%%%%%%%%%%%%%%%%%%%%%%%%%%
\begin{abstract}
The cost of the power distribution infrastructures is driven by the peak power encountered in the system. Therefore, the distribution network operators consider billing consumers behind a common transformer in the function of their peak demand and leave it to the consumers to manage their collective costs. This management problem is, however, not trivial. In this paper, we consider a multi-agent residential smart grid system, where each agent has local renewable energy production and energy storage, and all agents are connected to a local transformer. The objective is to develop an optimal policy that minimizes the economic cost consisting of both the spot-market cost for each consumer and their collective peak-power cost. We propose to use a parametric Model Predictive Control (MPC)-scheme to approximate the optimal policy. The optimality of this policy is limited by its finite horizon and inaccurate forecasts of the local power production-consumption. A Deterministic Policy Gradient (DPG) method is deployed to adjust the MPC parameters and improve the policy. Our simulations show that the proposed MPC-based Reinforcement Learning (RL) method can effectively decrease the long-term economic cost for this smart grid problem.
\end{abstract}
%%%%%%%%%%%%%%%%%%%%%%%%%%%%%%%%%%%%%%%%%%%%%%%%%%%%%%%%%%%%%%%%%%%%%%%%%%%%%%%%
\section{Introduction}
Nowadays, an increasing number of residents install and control their own small-scale energy system distributed throughout the whole grid \cite{mishra2013scaling}. A typical multi-agent residential smart grid system is illustrated in Fig. \ref{fig:grid}. In each ``agent" as labeled in Fig. \ref{fig:grid}, the power consumption comes from the daily life usage, Electrical Vehicle (EV) charger, etc. Home-scale renewable energies, like solar Photo Voltaic (PV), are used to generate electricity. Batteries are the most widely used devices for storing excess power. The spot-market prices (hourly prices announced at least 11-12$\si{h}$ ahead) incentivize residential prosumers to consume and/or store energy at low prices and consume less and/or use the stored energy to meet demands when prices are high \cite{hegde2011optimal,albadi2008summary}. However, the spot market does little to reduce the peak power demand. If all the houses in a neighborhood tend to require power exchanges with the power grid simultaneously, then the local transformer must be sized accordingly, incurring high investment costs for the distribution system. With the electrification of our society, this issue is becoming more severe. The power distribution companies are thus trying to reduce the peak power consumers impose on the system via some economic incentives. An ideal economic incentive structure is to collectively invoice the consumers behind a given transformer for its monthly peak power. This best reflects the peak-power cost on the distribution infrastructures \cite{fang2011smart,madlener2002power}. In this scenario, the goal is to find an optimal smart-grid policy that minimizes the spot-market cost for each agent while decreasing the collective peak-power cost for the whole system. 

\par Economic Model Predictive Control (EMPC) has been the natural choice to solve this kind of problem, see for instance \cite{cole2014optimal,knudsen2017model,godina2018model,dumas2021coordination}. However, setting up an EMPC-scheme that delivers an optimal policy remains challenging because of the stochastic nature of the local power production-consumption \cite{koutsopoulos2011optimal}, and the relatively short EMPC prediction horizon (up to $24\si{h}$, i.e. the period for which the spot market prices are available~\cite{price}) compared to the peak-power billing period of one month. Therefore, the policy resulting from an ordinary EMPC-scheme is prone to be suboptimal and could be further improved.

\par Recently, Reinforcement Learning (RL)-based control strategies are gaining attention, as they can make good use of data to reduce the impact of uncertainties and disturbances \cite{zhang2018review, remani2018residential}, without relying on a model that captures the real system accurately. However, regular RL would arguably be a poor choice for the problem considered here due to the tremendous amount of data required to learn an optimal policy from scratch. Besides, the presence of local production-consumption forecasts in the form of time-series in the decision policies creates high-dimensional information spaces that are detrimental to fast model-free learning. 

\par RL typically relies on Deep Neural Networks (DNNs) as function approximators \cite{bucsoniu2018reinforcement}. Nevertheless, DNN-based RL does not provide formal tools to discuss the closed-loop stability and constraints satisfaction. Moreover, selecting the initial weights of a DNN-based policy is difficult and often done randomly, which leads to a lengthy learning process. In contrast, MPC-based policies benefit from a large set of theoretical tools addressing those issues \cite{zanon2020safe}, and could provide fairly effective policies by using the available system models. Therefore, we propose to combine the MPC strategy with RL: using a parametrized MPC-scheme as the function approximation of the optimal policy and use RL to adjust the parameters such that the system performance is optimized. The combination of MPC and RL has been proposed and justified in~\cite{zanon2020safe,gros2019data}, where it is shown that an EMPC scheme can theoretically generate the optimal policy for a given system even if the EMPC model is inaccurate. Recent researches have further developed and demonstrated this approach~\cite{Arash2021CCTAMPC,gros2020reinforcement,zanon2019practical,Wenqi2021CDC}.

\par In this paper, the MPC-based RL method is adopted to seek an optimal smart-grid policy that minimizes the long-term economic costs, including the spot-market cost and the peak-power cost. We use a parametrized MPC-scheme to approximate the optimal policy suffering from varying spot-market prices and inaccurate local agent's power production-consumption forecasts. To improve the closed-loop performance of the MPC-based policy, a Deterministic Policy Gradient (DPG) method is deployed based on the actor-critic setting. The critic is computed using a Least Squares Temporal Difference (LSTD) method; The MPC (i.e., the actor) parameters are updated using the policy gradient, bringing the MPC-based policy gradually approach to the (sub)optimal policy.
%%%%%%%%%%%%%%%%%%%%%%%%%%%%%%%%%%%%%%%%%%%%%%%%%%%%%%%%%%%%%%%%%%%%%%%%%%%%%%%%
\section{Problem Formulation}
This section introduces the system dynamics and the objective function for this smart grid problem.
\subsection{System}
\begin{figure}[htbp!]
\centering
\includegraphics[width=0.4\textwidth]{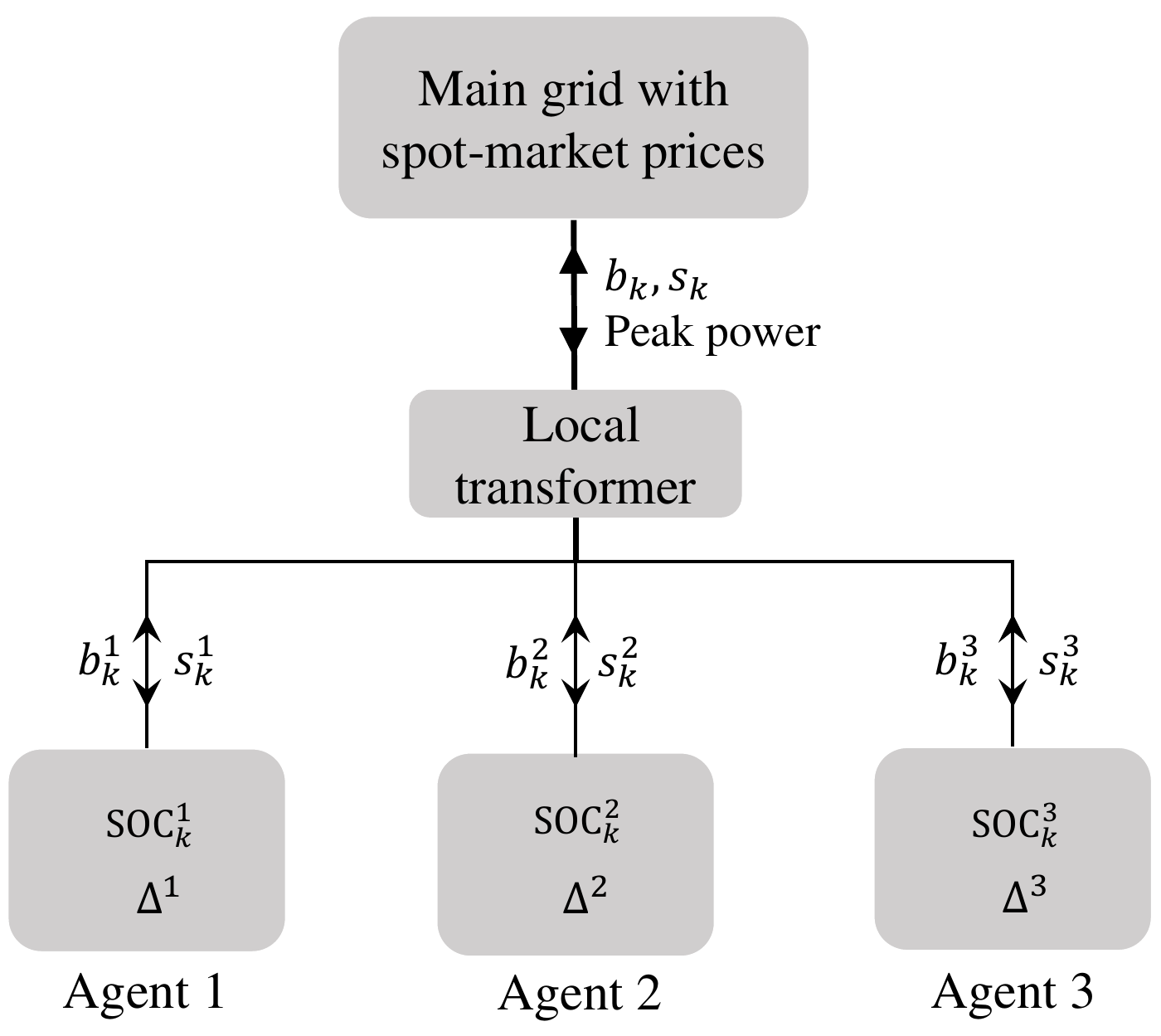}
\caption{Illustration of a typical three-agent residential smart grid system.
}
\label{fig:grid}
\end{figure}

Consider a real-world residential smart grid system as shown in Fig. \ref{fig:grid}. The dynamics of the agent is modelled as
\begin{align} 
\mathrm{soc}^i_{k+1} &= \mathrm{soc}^i_{k} + \alpha^i\left(\Delta^i_k + b^i_k- s^i_k \right)\label{eq:Dyna00} \\ \Delta^i_{k+1}&=\beta^i\Delta^i_k+\delta^i_k ,
\label{eq:Dyna01}
\end{align} 
where superscript $i=1,\ldots, N_a$ denotes the index of $i^{\si{th}}$ agent among the total $N_a$ agents. The sampling time is $1\si{h}$ as the spot-market price is hourly and the subscript $k=0,1,2\ldots$ is the physical time index. Scalar $\mathrm{soc}^i_k\in[0,1]$ is a dimensionless parameter corresponding to the relative State-of-Charge (SOC) of agent $i$ at time $k$. The interval $[0,1]$ represents the SOC level considered as non-damaging for the battery (typically $20\%-80\%$ range of the physical SOC). Constant $\alpha^i \left[{\si{kWh^{-1}}}\right]>0$ stands for the battery size. Input $b^i_k$ (resp. $s^i_k$) $\left[\si{kWh}\right]$ $\in[0,\bar U^i]$ is the energy bought (resp. sold) from (resp. to) the main grid in the time interval $[t_k,t_{k+1}]$, where $\bar U^i$ is the maximum allowed buying (resp. selling) electric quantity of the $i^{\mathrm{th}}$ battery. State $\Delta^i_k\left[\si{kWh}\right]$ is the difference between the local power production and consumption in the sampling interval $[t_k,t_{k+1}]$. Forecasts for $\Delta^i_k$ are typically available. For the sake of simplicity, here we consider that $\Delta_k^i$ is a stable random walk \cite{le1991range} averaging at zero. Constant $\beta^i\in[0,1)$ is the stabilizing coefficient. Normal centred random variable $\delta_k^i\sim\mathcal N\left( \mu^i, \sigma^i\right)$ introduces the stochasticity, where $\mu^i$ and $\sigma^i$ are the mean and variance of the Normal distribution. Besides, note that here we assume the battery efficiency of $100\%$ during the charging/discharging process.
%%%%%%%%%%%%%%%%%%%%%%%%%%%%%%%%%%%%%%%%%%%%%%%%%%%%%%%%%%%%%%%%%%%%%%%%%%%%%%%%
\subsection{Cost function}
\par The spot-market cost is assumed to be linearly related to the difference between the energy bought from and sold to the grid \cite{harsha2014optimal}. Hence, each agent has the following spot-market cost
\begin{align} 
\label{eq:Le} L^i_{\mathrm{E}}(b^i_k,s^i_k) = \phi^b_k b^i_k-\phi^s_k s^i_k, 
\end{align}
where $\phi^b_k$ and $\phi^s_k$ are the buying and selling prices at time instance $t_k$. The spot-market stage cost $L_{\mathrm{E}}(\vect s_k,\vect a_k)$ for the entire system reads as the summation of each agent cost
\begin{align}
\label{eq:LE}
    L_{\mathrm{E}}(\vect s_k,\vect a_k)=\sum_{i=1}^{N_a} L^i_{\mathrm{E}}(b^i_k,s^i_k),
\end{align}
where vectors $\vect s_k=\{\mathrm{soc}_k^{1,\ldots,N_a},\Delta_k^{1,\ldots,N_a}\}$ and $\vect a_k=\{b_k^1,s_k^1,\ldots,b_k^{N_a},s_k^{N_a}\}$ are the system states and inputs \footnote{The state $\vect s_k$ and input $\vect a_k$ correspond to the state and action in the RL community.}, respectively.

\par As presented in Fig. \ref{fig:grid}, local agents collectively exchange power with the main grid via a common transformer, which is subject to the collective peak power of the agents connected to it. To reduce the peak power that consumers impose on the system, the ideal approach is to invoice the agents behind the transformer collectively for the monthly peak power seen by the transformer. 

\par To account for the peak-power cost in an EMPC setting, we first introduce the following monotonically non-decreasing sequence $P_{k}^{\mathrm{peak}}$
\begin{subequations}
\label{eq:ppeak} 
\begin{align} P_{k+1}^{\mathrm{peak}}&=\max\Bigg(P_{k}^{\mathrm{peak}}, \sum _{i=1}^{N_a} b^i_{k},\sum _{i=1}^{N_a}
s^i_{k}\Bigg),\\
P_{0}^{\mathrm{peak}}&=0.
\end{align} 
\end{subequations}
Variable $P_{k+1}^{\mathrm{peak}}$ describes the peak power up to $t_{k}$, and thus $P_{K+1}^{\mathrm{peak}}$ is the peak power for the entire interval $[t_0,t_K]$, where $t_K$ is the terminal time in the episodic task (assumed one month in this work with $t_K=720\si{h}$), i.e. we have
\begin{align}     
\label{eq:peak_power}
\text{Peak power in $[t_0,t_K]$ }:=P_{K+1}^{\mathrm{peak}}.
\end{align}
The peak-power cost is defined as a linear function of the peak power with a positive penalty coefficient $\lambda$, i.e.,
\begin{align}     
\label{eq:peak_power_cost}
\text{Peak-power cost}:=\lambda P_{K+1}^{\mathrm{peak}}.
\end{align}

\par Therefore, the objective is to find a control policy $\vect\pi$ that minimizes the monthly economic cost for the multi-agent system, expressed as
\begin{align}
\label{eq:J1} 
J(\vect\pi)=\mathbb E_{\vect\pi}\Bigg[\lambda P_{K+1}^{\mathrm{peak}}+\sum_{k=0}^{K} {L}_{{E}}(\vect s_k,\vect a_k) \Bigg|\vect a_k=\vect \pi(\vect s_k)\Bigg],
\end{align}
where the expectation $\mathbb E_{\vect\pi}$ is taken over the distribution of the Markov chain under the policy $\vect\pi$. To facilitate the deployment of the RL approach, we will recast the terminal peak-power cost \eqref{eq:peak_power_cost} as a stage cost. To this end, we define the peak-power stage cost $L_{P}(\vect s_k,\vect a_k)$ as follows
\begin{align}
\label{eq:peak_stage_cost}
L_{{P}}(\vect s_k,\vect a_k)=\lambda(P_{k+1}^{\mathrm{peak}}-P_{k}^{\mathrm{peak}}).
\end{align}
Obviously, \eqref{eq:peak_power_cost} equals to the sum of \eqref{eq:peak_stage_cost} within $[t_0,t_K]$, since
\begin{align}
    \sum_{k=0}^{K} {L_P}(\vect s_k,\vect a_k)&=\sum_{k=0}^{K}\lambda(P_{k+1}^{\mathrm{peak}}-P_{k}^{\mathrm{peak}})\nonumber\\
    &=\lambda(P_{K+1}^{\mathrm{peak}}-P_{0}^{\mathrm{peak}})\nonumber
    \\&=\lambda P_{K+1}^{\mathrm{peak}}.
\end{align}

\par Using \eqref{eq:LE} and \eqref{eq:peak_stage_cost}, we obtain the stage cost function $L(\vect s_k,\vect a_k)$ for this problem that consists of two terms (spot-market stage cost $L_{{E}}$ and peak-power stage cost $L_{{P}}$), written as
\begin{align} 
\label{eq:L}
L(\vect s_k, \vect a_k)=\underbrace{L_{\mathrm{E}}(\vect s_k,\vect a_k)}_{\text{Spot-market stage cost}} + \underbrace{L_{{P}}(\vect s_k,\vect a_k)}_{\text{Peak-power stage cost}}.
\end{align}
Consequently, the \emph{closed-loop performance} $J(\vect\pi)$ \eqref{eq:J1} of the whole system could be equivalently expressed as the sum of the stage cost \eqref{eq:L} over interval $[t_0,t_K]$, i.e.,
\begin{align}
\label{eq:J}
J(\vect\pi)=\mathbb E_{\vect\pi}\Bigg[&\sum_{k=0}^{K} {L}(\vect s_k,\vect a_k) \Bigg|\vect a_k=\vect \pi(\vect s_k)\Bigg].
\end{align}
Note that no discount factor is used here as we consider an episodic scenario over the monthly billing period.
%%%%%%%%%%%%%%%%%%%%%%%%%%%%%%%%%%%%%%%%%%%%%%%%%%%%%%%%%%%%%%%%%%%%%%%%%%%%%%%%
\section{MPC as an optimal policy approximator}
\label{sec:MPC}
For this smart grid problem, EMPC is a natural tool to support the optimal policy approximation. However, since the EMPC horizon ($12\si{h}$) is much shorter than the episode length ($720\si{h}$), and the local power production-consumption forecasting model is imprecise, the policy generated by the ordinary EMPC scheme would be significantly suboptimal. Therefore, it is sensible to parameterize the EMPC scheme by adding parameters in the model, cost, and constraints of the MPC scheme, and let RL adjust these parameters to minimize the closed-loop performance of the system. In this section, the parameterized MPC-scheme as a function approximator of the optimal smart-grid policy $\vect\pi^\star$ is presented.

\par Consider the following MPC-scheme parametrized with $\vect \theta$
\begin{subequations}
\label{eq:mpc}
\begin{align}
    &\min_{\widehat {\mathrm{\vect {soc}}},\hat {\vect \Delta}, \hat {\vect b},\hat {\vect s}}\,\,\,\, \theta_{\lambda}\hat P^{\mathrm{peak}}_N+\sum_{i=1}^{N_a}\Bigg(  T(\widehat{\mathrm{soc}}^i_{N},\theta^i_T,\theta^i_{\varphi})+\label{eq:mpc_cost} \\&\null\qquad\qquad\qquad \sum_{j=0}^{N-1}\left(  L_{E}(\hat{b}^i_j,
    \hat s^i_j)+\psi(\widehat{\mathrm{soc}}^i_{j},\theta^i_\psi,\theta^i_{\varphi})\right)\Bigg)\nonumber \\
    &\quad\quad\mathrm{s.t.}\quad
    \forall i=1,\ldots,N_a , \quad\forall j=0,\ldots,N-1\nonumber
    \\&\null \,\,\qquad\qquad \widehat{\mathrm{soc}}^i_{j+1} = \widehat{\mathrm{soc}}^i_{j} + \theta_{\alpha}^i(\hat\Delta^i_k+\hat b^i_j-\hat s^i_j), \label{eq:MPC:dynamics}\\
    &\null \,\,\qquad\qquad \hat\Delta^i_{j+1}=\theta_\beta^i\hat\Delta^i_j+\theta_\delta^i, \label{eq:MPC:dynamics1}\\
    &\null \,\,\qquad\qquad 0\leq\widehat{\mathrm{soc}}^i_{j}\leq 1,\qquad 0\leq\widehat{\mathrm{soc}}^i_{N}\leq 1, \label{eq:MPC:state:cons1}\\
    &\null\,\,\qquad\qquad 0\leq \hat b^i_j\leq \bar U^i,\qquad\,\,
0\leq \hat s^i_j\leq \bar U^i,\label{eq:MPC:input:cons}\\ &\null\,\,\qquad\qquad  \hat P^{\mathrm{peak}}_{j+1}=\max\left(\hat P^{\mathrm{peak}}_{j}, \sum _{i=1}^{N_a} \hat b^i_{j},\sum _{i=1}^{N_a} \hat s^i_{j}\right), \label{eq:MPC:equality}\\&\null\,\,\qquad\qquad \widehat{\mathrm{soc}}^i _{0}=\mathrm{soc}^i _{k}, \,\hat { \Delta}^{i}_{0} = \Delta^{i}_{k},\,\hat P^{\mathrm{peak}}_{0} = P^{\mathrm{peak}}_{k}, \label{eq:MPC:initial}
\end{align}
\end{subequations}
where $N$ is the prediction horizon. Arguments $\widehat{\vect{\mathrm{soc}}}=\{\widehat{\mathrm{soc}}^{1,\dots,N_{a}}_{0,\dots,N}\}$, $\hat {\vect \Delta}=\{\hat { \Delta}^{1,\dots,N_{a}}_{0,\dots,N}\}$, $\hat{\vect{b}}=\{\hat{b}^{1,\dots,N_{a}}_{0,\dots,N-1}\}$, and $\hat{\vect{s}}=\{\hat{s}^{1,\dots,N_{a}}_{0,\dots,N-1}\}$ are the primal decision variables. Formula \eqref{eq:mpc_cost} is the MPC cost, including the spot-market stage cost $L_E(\cdot)$ and the peak-power cost $\theta_{\lambda}\hat P^{\mathrm{peak}}_N$. The additional stage cost $\psi(\cdot)$ and terminal cost $T(\cdot)$ are introduced to serve as cost modifiers, as detailed in \cite{gros2019data}. Note that functions $\psi(\cdot),T(\cdot)$ and the peak-power cost function are all parametrized by $\vect \theta$. Equations \eqref{eq:MPC:dynamics}, \eqref{eq:MPC:dynamics1} describe the parameterized system dynamics. Equations \eqref{eq:MPC:state:cons1}, \eqref{eq:MPC:input:cons} introduce the states/inputs restrictions. The peak power condition and the MPC initialization are handled in \eqref{eq:MPC:equality} and \eqref{eq:MPC:initial}, respectively.

\par The parameters $\vect \theta$ are gathered as
\begin{align}    
\label{eq:theta}
\vect\theta:=\{\theta^i_{\alpha},\theta^i_{\beta},\theta^i_{\delta},\theta^i_{T},\theta^i_{\psi},\theta^i_{\varphi},\theta_{\lambda}\},\quad i=1,\dots,N_{a}
\end{align}  
in which $\theta^i_{\alpha}$ is designed to compensate for the model uncertainties, parameters $\theta^i_{\beta}$ and $\theta^i_{\delta}$ coarsely model the stochasticities of the power production-consumption forecasts. $\theta^i_{T}$ (with $\theta^i_{\varphi}$) shapes the terminal cost function of the EMPC scheme. The value of the terminal cost is important here as the peak-power cost \eqref{eq:peak_power_cost} pertains to the entire billing period, which is much longer than the short EMPC horizon. The ideal adjustment of $\theta^i_{T}$ in the terminal cost is difficult to perform manually and is hence left to RL. Parameter $\theta^i_{\psi}$ (with $\theta^i_{\varphi}$) can modify the stage cost along the lines of \cite{gros2019data}, and $\theta^i_{\varphi}$ allows RL to assign some preferred SOC levels in the EMPC scheme in view of minimizing its long-term performance. Parameter $\theta_{\lambda}$ weighs the peak-power cost against the spot-market cost. A larger $\theta_{\lambda}$ implies a greater focus on the peak-power cost in the total cost. All these parameters are adjusted via RL in the direction that minimizes the long-term economic cost \eqref{eq:J}. It is worth mentioning that this choice of parameterization is arbitrary and different options are possible. Theoretically, under some assumptions, if the parametrization is rich enough, the MPC scheme is capable of capturing the optimal policy $\vect\pi^\star$ in the presence of uncertainties and disturbances \cite{gros2019data}.

\par The parameterized policy for agent $i$ at time $k$ is obtained based on applying the first input of the input sequence delivered by the MPC \eqref{eq:mpc}, i.e.,
\begin{align}
     \vect\pi_{\vect\theta}^{i}(\vect s_k)= \left[\hat b_0^{i\star}(\vect s_k,\vect \theta), \hat s_0^{i\star}(\vect s_k,\vect \theta)\right]^\top,
\end{align}
 where $\hat b_0^{i\star}$ and $\hat s_0^{i\star}$ are the first elements of $\hat {\vect{b}}^{i\star}$ and $\hat {\vect{s}}^{i\star}$, which are the solutions of the MPC scheme \eqref{eq:mpc} associated to the decision variables $\hat {\vect{b}}^{i}$ and $\hat {\vect{s}}^{i}$. Then, the global parametric smart-grid policy could be written as
 \begin{align}\label{eq:pi}
  \vect\pi_{\vect\theta}(\vect s_k)=\left[{\pi_{\vect\theta}^1}^\top,\ldots,{\pi_{\vect\theta}^{N_a}}^\top\right]^\top.   
 \end{align}
Besides, note that the input (action) $\vect a_k$ in RL is selected according to the parametric policy $\vect\pi_{\vect\theta}$ in \eqref{eq:pi} with some small random exploration.
%%%%%%%%%%%%%%%%%%%%%%%%%%%%%%%%%%%%%%%%%%%%%%%%%%%%%%%%%%%%%%%%%%%%%%%%%%%%%%%%%
\section{Deterministic Policy Gradient Method}
This section details how the DPG method adjusts the parameters \eqref{eq:theta} in the MPC scheme \eqref{eq:mpc}. DPG method \cite{SuttonPG, sutton2018reinforcement}, as a direct RL approach, optimizes the policy parameters $\vect\theta$ directly via gradient descent steps on the performance function $J$ \eqref{eq:J}. The update rule is as follows
\begin{align}
\label{eq:update theta}
    \vect\theta \leftarrow \vect\theta-\eta  \nabla _{\vect\theta}J(\vect\pi _{\vect\theta}),
\end{align}
where $\eta>0$ is the step size, and the gradient of $J$ with respect to parameters $\vect\theta$ is obtained as \cite{silver2014deterministic}
\begin{align}\label{eq:dj}
    \nabla _{\vect\theta}J(\vect\pi _{\vect\theta}) = \mathbb E\left[{\nabla _{\vect\theta} }{\vect\pi _{\vect\theta} }(\vect s){\nabla _{\vect a}}{Q_{{\vect\pi _{\vect\theta} }}}(\vect s,\vect a)|_{\vect a=\vect \pi _{\vect\theta}}\right],
\end{align}
where $Q_{\vect{\pi}_{\vect{\theta}}}$ is the action-value function associated to the policy ${\vect{ \pi}}_{\vect{\theta}}$. The calculations of ${\nabla _{\vect\theta} }{\vect\pi _{\vect\theta} }(\vect s)$ and ${\nabla _{\vect a}}{Q_{{\vect\pi _{\vect\theta} }}}(\vect s,\vect a)$ in (\ref{eq:dj}) are discussed in the following.
%%%%%%%%%%%%%%%%%%%%%%%%%%%%%%%%%
\subsubsection{${\nabla _{\vect\theta} }{\vect\pi _{\vect\theta} }(\vect s)$}
The primal-dual Karush Kuhn Tucker (KKT) conditions underlying the MPC scheme \eqref{eq:mpc} is written as
\begin{align}
    \vect R = {\left[ {\begin{array}{*{20}{c}}
{{\nabla _{\vect \zeta}}{\mathcal L_{\vect\theta} }}^\top&{{\vect G_{\vect\theta} }}^\top &{\vect H_{\vect\theta}^\top\mathrm{diag}\left(\vect\mu\right)  }
\end{array}} \right]^\top},
\end{align}
where $\vect\zeta=\{\hat {\mathrm{\vect {soc}}},\hat{\vect\Delta},\hat {\vect b},\hat {\vect s}\}$ is the primal decision variable of the MPC \eqref{eq:mpc}. Operator ``$\mathrm{diag}$" assigns the vector elements onto the diagonal elements of a square matrix. $\mathcal{L}_{\vect \theta}$ is the Lagrange function associated to \eqref{eq:mpc}, written as
\begin{align}
\mathcal{L}_{\vect \theta}(\vect y) = \Omega_{\vect \theta} + \vect\lambda^\top \vect G_{\vect\theta}  + \vect\mu^\top \vect H_{\vect \theta},
\end{align}
where $\Omega_{\vect\theta}$ is the MPC cost \eqref{eq:mpc_cost}, $\vect G_{\vect\theta}$ gathers the equality constraints and $\vect H_{\vect\theta}$ collects the inequality constraints of the MPC \eqref{eq:mpc}. Vectors $\vect\lambda,\vect\mu$ are the associated dual variables. Argument ${\vect y}$ reads as ${\vect y} =\{\vect\zeta,\vect\lambda,\vect\mu\}$ and $ {\vect y}^{\star}$ refers to the solution of the MPC \eqref{eq:mpc}. Consequently, the policy sensitivity ${\nabla _{\vect \theta} }{\vect \pi _{\vect \theta} }$ required in \eqref{eq:dj} can be obtained as follows~\cite{gros2019data}
\begin{align}
\label{eq:sensetivity}
{\nabla _{\vect \theta} }{\vect \pi _{\vect \theta} }\left(\vect  s \right) =  - {\nabla _{\vect\theta} }{\vect R }\left( {\vect y^\star},\vect s,\vect\theta\right){\nabla _{\vect y}}{\vect R }{\left( {\vect y^\star},\vect s,\vect\theta \right)^{ - 1}}\frac{\partial {\vect y}}{\partial {\vect u_0}},
\end{align}
where $\vect u_0$ is the first element of the input sequence.
%%%%%%%%%%%%%%%%%%%%%%%%%%%%%%%%%%%%%%%%%
\subsubsection{${\nabla _{\vect a}}{Q_{{\vect\pi _{\vect\theta} }}}(\vect s,\vect a)$}
Under some conditions~\cite{silver2014deterministic}, the action-value function $Q_{{\vect\pi _{\vect\theta} }}$ can be replaced by an approximator ${Q_{\vect w}}$, i.e. $Q_{\vect w}\approx Q_{\vect \pi_{\vect \theta}}$, without affecting the policy gradient. Such an approximation is labelled \textit{compatible} and can, e.g., take the form
\begin{align}
\label{eq:Q_w}
{Q_{\vect w}}\left( {\vect s,\vect a} \right) = {{\left( {\vect a - {\vect\pi _{\vect\theta} }\left( \vect s \right)} \right)}^{\top}}{\nabla _{\vect\theta} }{\vect\pi _{\vect\theta} }{{\left( \vect s \right)}^{\top}}\vect w + { V_{\vect v}}\left( \vect s \right),
\end{align}
where $\vect w$ is a set of parameters supporting the approximation $Q_{\vect w}$ of the action-value function $Q_{{\vect{ \pi}}_{\vect{\theta}}}$ and $V_{\vect v}\approx V_{\vect \pi_{\vect \theta}}$ is the parameterized baseline function approximating the value function. It can take a linear form
\begin{align}
\label{eq:V_v}
    {V_{\vect v}} \left(\vect s\right ) =\vect \Phi\left(\vect s \right)^\top {\vect v},
\end{align}
where $\vect \Phi(\vect s)$ is the state feature vector and $\vect v$ is the corresponding parameters vector. Using \eqref{eq:Q_w}, we obtain
\begin{align}
    {\nabla _{\vect a}}{Q_{{\vect\pi _{\vect\theta} }}}(\vect s,\vect a) \approx {\nabla _{\vect a}}{Q_{\vect w}}(\vect s,\vect a)=\nabla _{\vect\theta}\vect \pi _{\vect\theta}\left(\vect s\right )^{\top}\vect w.
\end{align}
The parameters $\vect w$ and $\vect v$ of the action-value function approximation \eqref{eq:Q_w} ought to be computed as the solution of the Least Squares (LS) problem
\begin{align}
\label{eq:error}
    \min_{\vect w, \vect v} \mathbb{E} \left[\big( Q_{\vect\pi_{\vect\theta}}(\vect s,\vect a)-Q_{\vect w} (\vect s,\vect a)\big )^2\right],
\end{align}
which, in this work, is tackled via the Least Squares Temporal Difference (LSTD) method (see \cite{lagoudakis2003least}). Finally, equation \eqref{eq:update theta} can be rewritten as a compatible DPG
\begin{align}
\label{eq:new_theta}
    {\vect \theta} \leftarrow {\vect\theta} - \eta \mathbb E \left[ {\nabla _{\vect\theta} }{\vect\pi _{\vect\theta} }\left( \vect s_k \right) {{\nabla _{\vect\theta} }{\vect\pi _{\vect\theta} }{{\left( \vect s_k \right)}^{\top}}{\vect w}} \right].
\end{align}
%%%%%%%%%%%%%%%%%%%%%%%%%%%%%%%%%%%%%%%%%%%%%%%%%%%%%%%%%
\section{Simulation}
This section provides the simulation results of the proposed MPC-based DPG method for a three-agent residential smart grid system. We consider different battery sizes $\alpha^{1,2,3}$ and stochastic local production-consumption $\Delta^{1,2,3}$, and apply the proposed MPC-based DPG approach to optimize the EMPC policy. The additional stage cost function $\psi(\cdot)$ and terminal cost function $T(\cdot)$ in \eqref{eq:mpc} are designed as
\begin{subequations}
\label{eq:T_psi}
\begin{align}     
\psi(\widehat{\mathrm{soc}}^i_{j},\theta_{\psi}^i,\theta^i_{\varphi})&=\theta_{\psi}^i(\widehat{\mathrm{soc}}^i_{j}-\theta^i_{\varphi})^2\\
T(\widehat{\mathrm{soc}}^i_{N},\theta^i_T,\theta^i_{\varphi})&=\theta_{T}^i(\widehat{\mathrm{soc}}^i_{N}-\theta^i_{\varphi})^2,
\end{align} 
\end{subequations}
which are quadratic functions of SOC with a adjustable parameter $\theta^i_{\varphi}$ as their setpoint and positive coefficient variables $\theta_{T}^i$ and $\theta_{\psi}^i$. One can use more generic function approximators in \eqref{eq:T_psi}, however, it is shown \cite{Arash2021MPC} that the quadratic parameterizations of $\psi(\cdot)$ and $T(\cdot)$ for this smart-grid problem are rich enough to capture the optimal policy. The spot-market buying prices $\phi^b$ are collected from the website provided by the Nord Pool European Power Exchange~\cite{price}, and the selling prices $\phi^s$ are assumed as half of the buying prices, i.e. $\phi^s=0.5\phi^b$ at every time step. The initial value of the parameter $\vect\theta$ is set as $\vect\theta_0$$=$$\left\{[0.01,0.012,0.014]^\top,\vect{0.9},\vect{0},\vect{20},\vect{20},\vect{0.5},1\right\}$, where the bold numbers represent constant vectors with suitable dimensions. Other parameters values used in the simulation are listed in Table \ref{tab:value}. 
%%%%%%%%%%%%
\begin{table}[htbp!]
\caption{Parameters values.}
\label{tab:value}
\centering
\begin{tabular}{cc}
\hline
symbol & \multicolumn{1}{|c}{value} \\
\hline
Sampling time & \multicolumn{1}{|c}{$1\si{h}$} \\
$N_a$   &  \multicolumn{1}{|c}{$3$} \\
$\alpha^{1,2,3}$ & \multicolumn{1}{|c}{$0.01,0.012,0.014$}  \\
$N$  & \multicolumn{1}{|c}{12}\\
$\beta^i$& \multicolumn{1}{|c}{$0.9$} \\
$\delta^i$  & \multicolumn{1}{|c}{$\mathcal{N}(0,0.5)$}\\
$\bar U^i$& \multicolumn{1}{|c}{$15$} \\
$K$ & \multicolumn{1}{|c}{$720$}\\
$\mathrm{soc}_0^i,\Delta_0^i$& \multicolumn{1}{|c}{$0.01,0$} \\
$\lambda$ & \multicolumn{1}{|c}{$100$}\\
${\vect\theta}_0$& \multicolumn{1}{|c}{$\left\{[0.01,0.012,0.014]^\top,\vect{0.9},\vect{0},\vect{20},\vect{20},\vect{0.5},1\right\}$}\\
\hline
\end{tabular}
\end{table}
%%%%%%%%%%
\begin{figure}[htbp!] 
\centering 
\includegraphics[width=0.48\textwidth]{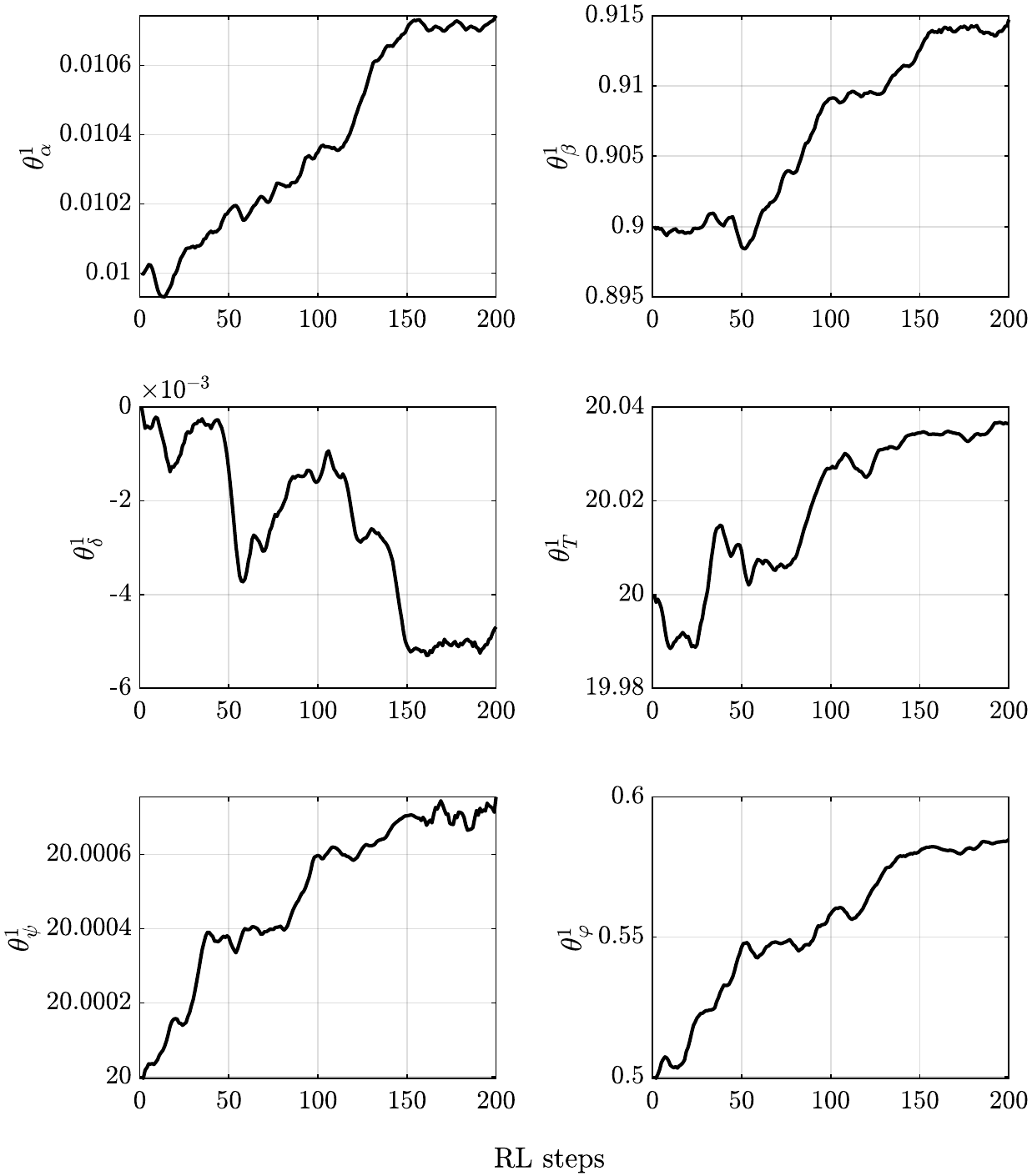} 
\caption{The variations of parameters $\{\theta^1_{\alpha},\theta^1_{\beta},\theta^1_{\delta},\theta^1_{T},\theta^1_{\psi},\theta^1_{\varphi}\}$ for agent $1$ over learning steps.}  
\label{fig:theta}
\end{figure}
%%%%%%%%
\begin{figure}[htbp!]
\centering
\includegraphics[width=0.35\textwidth]{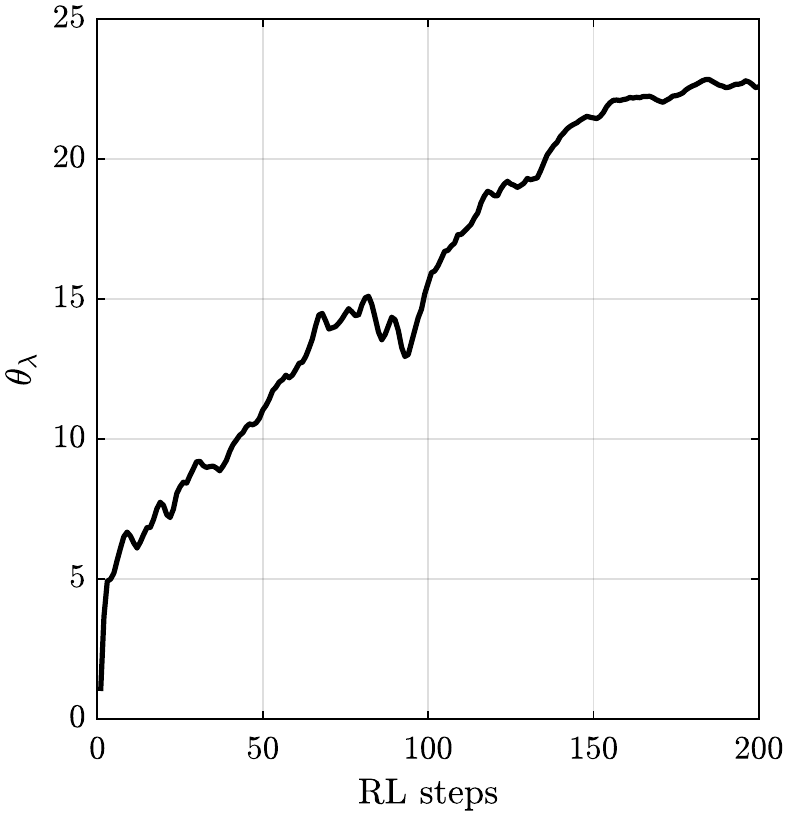}
\caption{The variation of parameter $\theta_{\lambda}$ over learning steps.}
\label{fig:lamda}
\end{figure}
%%%%%%%%
\begin{figure}[htbp!]
\centering
\includegraphics[width=0.35\textwidth]{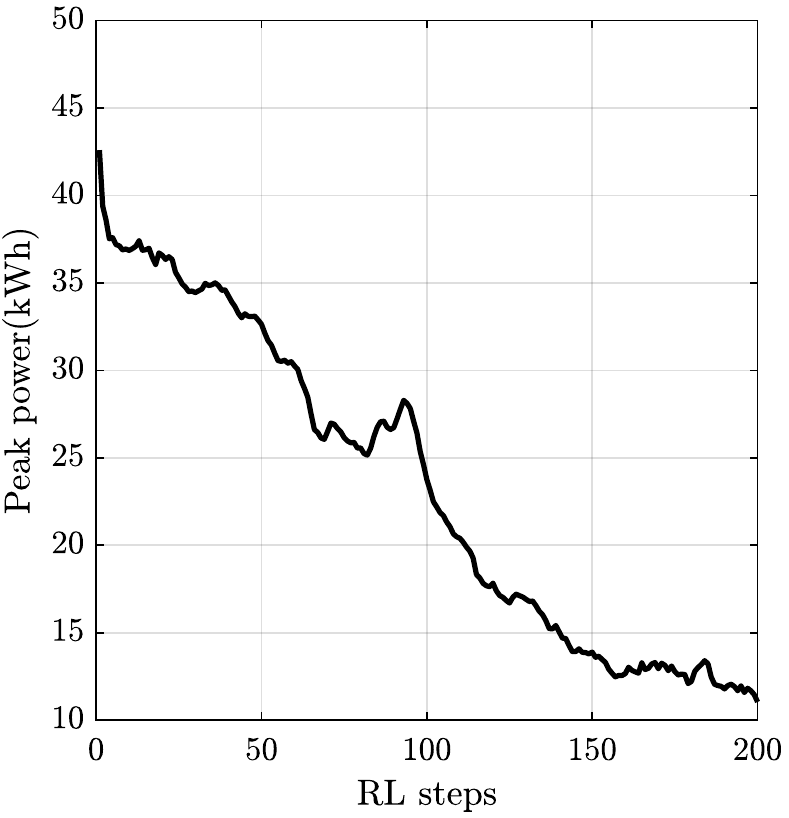}
\caption{The variation of the peak power over learning steps.}
\label{fig:peak}
\end{figure}
%%%%%%%%%%
\begin{figure}[htbp!] 
\centering 
\includegraphics[width=0.35\textwidth]{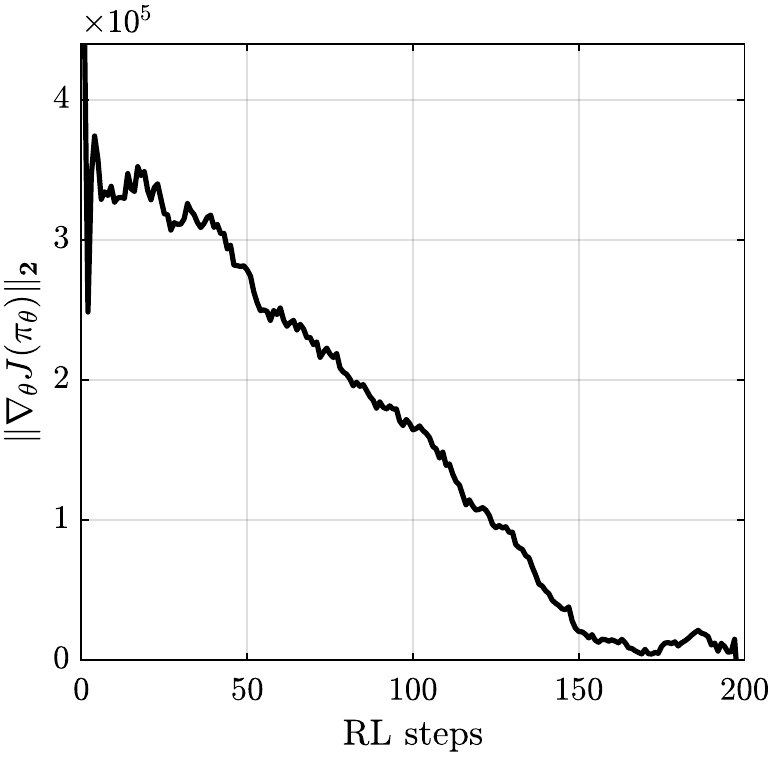}
\caption{The variation of the normed policy gradient $\|\nabla _{\vect\theta}J(\vect\pi _{\vect\theta})\|_2$ over learning steps.}  
\label{fig:J_gra}
\end{figure}
%%%%%%%%%%%%
\begin{figure}[htbp!] 
\centering 
\includegraphics[width=0.35\textwidth]{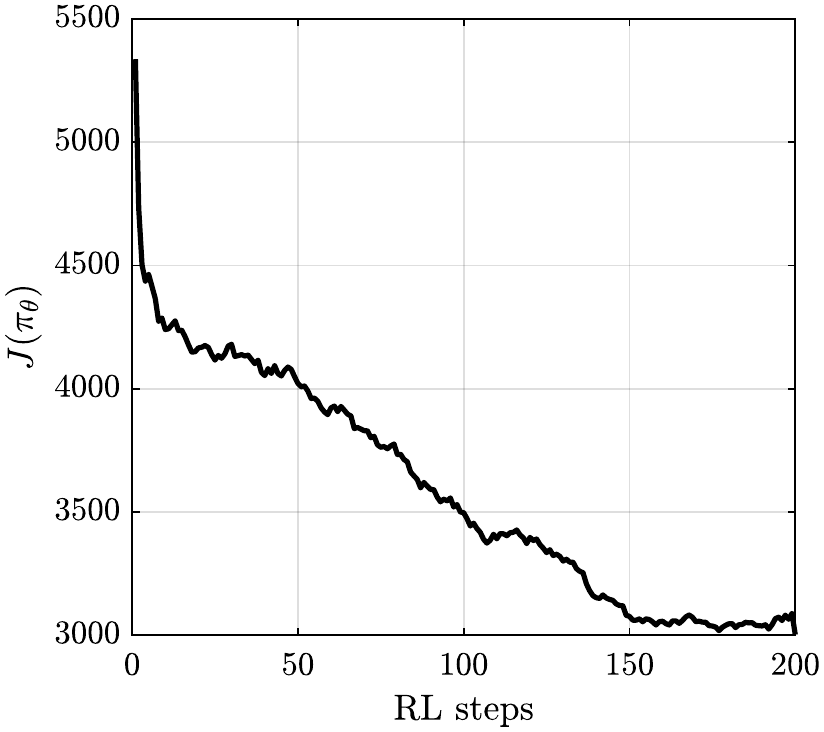}
\caption{The variation of the closed-loop performance $J(\vect\pi _{\vect\theta})$ over learning steps.}
\label{fig:J}
\end{figure}
%%%%%%%%%%%%
%%%%%%%%%%%%
\par Figure \ref{fig:theta} shows the parameters variations of agent $1$. Note that the variations of the parameters are different for the three agents, but for brevity, only the results of agent $1$ are presented representatively. $\theta^1_{\alpha},\theta^1_{\beta},\theta^1_{\delta}$ are initialized consistently with the real system parameters; $\theta^1_{T},\theta^1_{\psi}$ are set to start with $20$, taking into account of the proper weights of the parameterized cost functions; the initial value of $\theta^1_{\varphi}$ is selected as $0.5$ based on the assumption that batteries maintaining their SOC around half is preferable; nonetheless, the actual value should be determined by RL. It can be seen that all the parameters converge as learning progresses. 

\par Figure \ref{fig:lamda} provides the variation of the parameter $\theta_{\lambda}$ with respect to learning steps, which is the most important one among the $19$ parameters in the simulation. It can be seen that the value of $\theta_{\lambda}$ increases from $1$ and converges to around $22.5$, which signifies that in the MPC cost \eqref{eq:mpc_cost}, the concern on the peak-power cost is gradually increasing. As a result, the peak power should be reduced so as to minimize the peak-power cost. Besides, it is worth mentioning that the convergence value of $\theta_{\lambda}$ is $22.5$ instead of $100$ (the real value of the penalty factor $\lambda$ in \eqref{eq:peak_power_cost}). This is because, as we mentioned before, the actual RL peak-power cost is calculated for a period of one month, while MPC considers a much shorter interval of $12\si{h}$, so it is reasonable that the two values are not the same.

%%%%%%%%%%
%\begin{figure}[htbp!]
%\centering
%\includegraphics[width=0.48\textwidth]{soc.pdf}
%\caption{The $\mathrm{soc}^i_k$ of the three agents during episodes for the sampled RL steps: $1^{\mathrm{st}},50^{\mathrm{th}},100^{\mathrm{th}},150^{\mathrm{th}},200^{\mathrm{th}}$.}
%\label{fig:soc}
%\end{figure}
%%%%%%%%%%%%%%
%\par Figure \ref{fig:soc} shows the batteries SOC of each agent during episodes for five sampled RL steps: $1^{\mathrm{st}},50^{\mathrm{th}},100^{\mathrm{th}},150^{\mathrm{th}},200^{\mathrm{th}}$. It can be seen that the variances of $\mathrm{soc}^{1,2,3}$ decrease gradually with learning

%As can be observed, the purchases overwhelmingly occur when power prices are low, and when prices are high, the agents tend to use the stored energy from batteries and sell the excess power if have a surplus. This power-exchange strategy guarantees a low spot-market cost. Furthermore, from the beginning of learning (the $1^{\mathrm{st}}$ step) to the end of learning (the $200^{\mathrm{th}}$ step), the volume of buying/selling transactions for each agent decreases evidently due to the increase of $\theta_{\lambda}$. In other words, during prices troughs or peaks, the buying/selling strategies gradually tend to be more ``conservative" compared to the previous ``radical" ones, which avoids excessive-high peak power and therefore reduces the peak-power cost in \eqref{eq:J}. 
\par The peak power encountered in the system over learning steps is presented in Fig.~\ref{fig:peak}. As expected, it shows that the peak power continuously decreases with learning (from $42\si{kWh}$ down to $11\si{kWh}$), which corroborates the conclusions drawn in Fig.~\ref{fig:lamda}. The variation of the normed policy gradient $\|\nabla _{\vect\theta}J(\vect\pi _{\vect\theta})\|_2$ and the closed-loop performance $J(\vect \pi_{\vect \theta})$ are given in Fig.~\ref{fig:J_gra} and Fig.~\ref{fig:J}, respectively. It demonstrates that the policy gradient \eqref{eq:dj} is decreasing to zero with the learning proceeds. Correspondingly, the closed-loop performance \eqref{eq:J} decreases gradually and eventually reaches a (sub)optimum value, which indicates that we find the desired (sub)optimal policy that minimizes the long-term economic cost for this smart-grid problem.
%%%%%%%%%%%%%%%%%%%%%%%%%%%%%%%%%%%%%%%%%%%%%%%%%%%%%%%%%%%%%%%%%%%%%%%%%%%%%%%%%%%%%%%%%%%%%%%%%%%%%%%%%%%%%%%%%
\section{Conclusion}
In this paper, an MPC-based RL method is proposed to search for (sub)optimal smart-grid policies for multi-agent residential systems. Solving this problem is challenging due to the variability of power prices, the stochasticity of the agent's power production-consumption, and the ``long-term" nature of the problem. We show that our proposed approach could minimize the long-term economic costs that contain both the spot-market cost and the peak-power cost. Future works include the consideration of more sophisticated power management scenarios, where a distributed-MPC-based RL strategy would be a heuristic solution.
%%%%%%%%%%%%%%%%%%%%%%%%%%%%%%%%%%%%%%%%%%%%%%%%%%%%%%%%%%%%%%%%%%%%%%%%%%%%%%%%%%%%%%%%%%%%%%%%%%%%%%%%%%%%%%%%%
\bibliographystyle{IEEEtran}
\bibliography{References}

\begin{thebibliography}{10}
\providecommand{\url}[1]{#1}
\csname url@samestyle\endcsname
\providecommand{\newblock}{\relax}
\providecommand{\bibinfo}[2]{#2}
\providecommand{\BIBentrySTDinterwordspacing}{\spaceskip=0pt\relax}
\providecommand{\BIBentryALTinterwordstretchfactor}{4}
\providecommand{\BIBentryALTinterwordspacing}{\spaceskip=\fontdimen2\font plus
\BIBentryALTinterwordstretchfactor\fontdimen3\font minus
  \fontdimen4\font\relax}
\providecommand{\BIBforeignlanguage}[2]{{%
\expandafter\ifx\csname l@#1\endcsname\relax
\typeout{** WARNING: IEEEtran.bst: No hyphenation pattern has been}%
\typeout{** loaded for the language `#1'. Using the pattern for}%
\typeout{** the default language instead.}%
\else
\language=\csname l@#1\endcsname
\fi
#2}}
\providecommand{\BIBdecl}{\relax}
\BIBdecl

\bibitem{mishra2013scaling}
A.~Mishra, D.~Irwin, P.~Shenoy, and T.~Zhu, ``Scaling distributed energy
  storage for grid peak reduction,'' in \emph{Proceedings of the fourth
  international conference on Future energy systems}, 2013, pp. 3--14.

\bibitem{hegde2011optimal}
N.~Hegde, L.~Massouli{\'e}, T.~Salonidis \emph{et~al.}, ``Optimal control of
  residential energy storage under price fluctuations,'' \emph{Energy}, 2011.

\bibitem{albadi2008summary}
M.~H. Albadi and E.~F. El-Saadany, ``A summary of demand response in
  electricity markets,'' \emph{Electric power systems research}, vol.~78,
  no.~11, pp. 1989--1996, 2008.

\bibitem{fang2011smart}
X.~Fang, S.~Misra, G.~Xue, and D.~Yang, ``Smart grid—the new and improved
  power grid: A survey,'' \emph{IEEE communications surveys \& tutorials},
  vol.~14, no.~4, pp. 944--980, 2011.

\bibitem{madlener2002power}
R.~Madlener and M.~Kaufmann, ``Power exchange spot market trading in europe:
  theoretical considerations and empirical evidence,'' \emph{OSCOGEN
  (Optimisation of Cogeneration Systems in a Competitive Market
  Environment)-Project Deliverable}, vol.~5, 2002.

\bibitem{cole2014optimal}
W.~J. Cole, D.~P. Morton, and T.~F. Edgar, ``Optimal electricity rate
  structures for peak demand reduction using economic model predictive
  control,'' \emph{Journal of Process Control}, vol.~24, pp. 1311--1317, 2014.

\bibitem{knudsen2017model}
M.~D. Knudsen and S.~Petersen, ``Model predictive control for demand response
  of domestic hot water preparation in ultra-low temperature district heating
  systems,'' \emph{Energy and Buildings}, vol. 146, pp. 55--64, 2017.

\bibitem{godina2018model}
R.~Godina, E.~M. Rodrigues, E.~Pouresmaeil, J.~C. Matias, and J.~P.
  Catal{\~a}o, ``Model predictive control home energy management and
  optimization strategy with demand response,'' \emph{Applied Sciences},
  vol.~8, no.~3, p. 408, 2018.

\bibitem{dumas2021coordination}
J.~Dumas, S.~Dakir, C.~Liu, and B.~Corn{\'e}lusse, ``Coordination of
  operational planning and real-time optimization in microgrids,''
  \emph{Electric Power Systems Research}, vol. 190, p. 106634, 2021.

\bibitem{koutsopoulos2011optimal}
I.~Koutsopoulos, V.~Hatzi, and L.~Tassiulas, ``Optimal energy storage control
  policies for the smart power grid,'' in \emph{2011 IEEE International
  Conference on Smart Grid Communications (SmartGridComm)}.\hskip 1em plus
  0.5em minus 0.4em\relax IEEE, 2011, pp. 475--480.

\bibitem{price}
{Nord Pool Group}, ``Day-ahead power prices of {T}rondheim, {N}orway during
  {N}ovember, 2020,''
  \url{https://www.nordpoolgroup.com/Market-data1/Dayahead/Area-Prices/ALL1/Monthly/?view=table},
  2020.

\bibitem{zhang2018review}
D.~Zhang, X.~Han, and C.~Deng, ``Review on the research and practice of deep
  learning and reinforcement learning in smart grids,'' \emph{CSEE Journal of
  Power and Energy Systems}, vol.~4, no.~3, pp. 362--370, 2018.

\bibitem{remani2018residential}
T.~Remani, E.~Jasmin, and T.~I. Ahamed, ``Residential load scheduling with
  renewable generation in the smart grid: A reinforcement learning approach,''
  \emph{IEEE Systems Journal}, vol.~13, no.~3, pp. 3283--3294, 2018.

\bibitem{bucsoniu2018reinforcement}
L.~Bu{\c{s}}oniu, T.~de~Bruin, D.~Toli{\'c}, J.~Kober, and I.~Palunko,
  ``Reinforcement learning for control: Performance, stability, and deep
  approximators,'' \emph{Annual Reviews in Control}, vol.~46, pp. 8--28, 2018.

\bibitem{zanon2020safe}
M.~Zanon and S.~Gros, ``Safe reinforcement learning using robust {MPC},''
  \emph{IEEE Transactions on Automatic Control}, 2020.

\bibitem{gros2019data}
S.~Gros and M.~Zanon, ``Data-driven economic {NMPC} using reinforcement
  learning,'' \emph{IEEE Transactions on Automatic Control}, vol.~65, no.~2,
  pp. 636--648, 2019.

\bibitem{Arash2021CCTAMPC}
A.~B. Kordabad, W.~Cai, and S.~Gros, ``Multi-agent battery storage management
  using {MPC}-based reinforcement learning,'' \emph{arXiv preprint
  arXiv:2106.03541}, 2021.

\bibitem{gros2020reinforcement}
S.~Gros and M.~Zanon, ``Reinforcement learning for mixed-integer problems based
  on {MPC},'' \emph{arXiv preprint arXiv:2004.01430}, 2020.

\bibitem{zanon2019practical}
M.~Zanon, S.~Gros, and A.~Bemporad, ``Practical reinforcement learning of
  stabilizing economic {MPC},'' in \emph{2019 18th European Control Conference
  (ECC)}.\hskip 1em plus 0.5em minus 0.4em\relax IEEE, 2019, pp. 2258--2263.

\bibitem{Wenqi2021CDC}
W.~Cai, A.~B. Kordabad, H.~N. Esfahani, A.~M. Lekkas, and S.~Gros,
  ``{MPC}-based reinforcement learning for a simplified freight mission of
  autonomous surface vehicles,'' \emph{arXiv preprint arXiv:2106.08634}, 2021.

\bibitem{le1991range}
J.-F. Le~Gall and J.~Rosen, ``The range of stable random walks,'' \emph{The
  Annals of Probability}, pp. 650--705, 1991.

\bibitem{harsha2014optimal}
P.~Harsha and M.~Dahleh, ``Optimal management and sizing of energy storage
  under dynamic pricing for the efficient integration of renewable energy,''
  \emph{IEEE Transactions on Power Systems}, vol.~30, no.~3, pp. 1164--1181,
  2014.

\bibitem{SuttonPG}
R.~S. Sutton, D.~A. McAllester, S.~P. Singh, and Y.~Mansour, ``Policy gradient
  methods for reinforcement learning with function approximation,'' in
  \emph{Advances in neural information processing systems}, 2000, pp.
  1057--1063.

\bibitem{sutton2018reinforcement}
R.~S. Sutton and A.~G. Barto, \emph{Reinforcement learning: An
  introduction}.\hskip 1em plus 0.5em minus 0.4em\relax MIT press, 2018.

\bibitem{silver2014deterministic}
D.~Silver, G.~Lever, N.~Heess, T.~Degris, D.~Wierstra, and M.~Riedmiller,
  ``Deterministic policy gradient algorithms,'' in \emph{Proceedings of the
  31st International Conference on International Conference on Machine
  Learning}.\hskip 1em plus 0.5em minus 0.4em\relax JMLR.org, 2014, p.
  I–387–I–395.

\bibitem{lagoudakis2003least}
M.~G. Lagoudakis and R.~Parr, ``Least-squares policy iteration,'' \emph{Journal
  of machine learning research}, vol.~4, pp. 1107--1149, 2003.

\bibitem{Arash2021MPC}
A.~B. Kordabad, W.~Cai, and S.~Gros, ``{MPC}-based reinforcement learning for
  economic problems with application to battery storage,'' \emph{arXiv preprint
  arXiv:2104.02411}, 2021.

\end{thebibliography}
\end{document}